\documentclass[11pt,twoside]{article}
\usepackage{asp2014}

\aspSuppressVolSlug
\resetcounters

\begin{document}

\title{A Modelers' Opacity Wish List}
\author{Regner Trampedach$^{1,2,3}$\\
\affil{$^1$Space Science Institute, Boulder, CO 80301 USA; \email{rtrampedach@SpaceScience.org}}
\affil{$^2$Stellar Astrophysics Centre, Dept. of Physics and Astronomy, Aarhus University, DK--8000 Aarhus C, Denmark}
\affil{$^3$Lab. for Atmospheric and Space Physics, University of Colorado, Boulder, CO 80303, USA}
}

% This section is for ADS Processing.  There must be one line per author.
\paperauthor{Regner Trampedach}{rtrampedach@SpaceScience.org}{}{Space Science Institute}{}{Boulder}{CO}{80301}{USA}

%\aindex{Trampedach, R.}

\begin{abstract}
  At the Workshop on Astrophysical Opacities, several attendees voiced their interest in a list of absorption data that are missing from or inadequate in current models of astrophysical objects. This wish list by modelers is meant as motivation and inspiration for experimentalists and theoreticians alike.
\end{abstract}

\section{The Wish List}

Table~\ref{TheTable} is the opacity wish list as I have compiled from the responses received from both the attendees of the Workshop on Astrophysical Opacities and several 1D and 3D stellar atmosphere groups. The first column lists the absorber system. The second lists the absorption mechanism: rotational, vibrational, and electronic transitions in molecules; bound--bound (bb) and bound--free (bf) electronic transitions in atoms/ions; and collision induced absorption (CiA) from transient dipoles by pairs of passing particles. The third and fourth columns list the requests for laboratory experiment and/or theoretical calculations, and the last lists the kind of objects for which modeling will improve with updated atomic/molecular data.

\begin{table}[!ht]
  \caption{The opacity wish list}
  \label{TheTable}
  \smallskip
  {\centering\small
  \begin{tabular}{llccl}
  \tableline
  \noalign{\smallskip}
  Absorber & Mechanism & Exp. & Calc.& For objects \\
  \noalign{\smallskip}
  \tableline
  \noalign{\smallskip}
  CH$_2$, C$_3$H                      & Rot+vib+e & X & X & Cool star/C-star atm. \\
  CaOH, LaO, YO                       & Rot+vib+e & X & X & Cool dwarf atm.       \\
  ScO, TiS, ZrS                       & Rot+vib+e & X & X & Cool dwarf atm.       \\
  C$_2$H$^{\rm a}$                    & Rot+vib   &   & X & C-star atm.           \\
  N$_2$--H$_2$, N$_2$--He, N$_2$--Ne  & CiA       &   & X & M-dwarf atm.          \\
  N$_2$--N$_2$, H$_2$--Ne             & CiA       &   & X & M-dwarf atm.          \\
  H--H$^{\rm b}$                      & CiA       &   & X & Cool dwarf atm.       \\
  O, Fe-peak, Zn, Pb$^{\rm c}$        & bb, bf    & X & X & Stars in general      \\
  \noalign{\smallskip}
  \tableline
  \end{tabular}

  }
  \scriptsize
    \ \ \ \ \ \ \ \ \ \ \ \ \ {\bf Notes.} $^{\rm a}$No data are currently available. $^{\rm b}$This is a request for an update of the old \citet{doyle:H+H} result,

    \ \ \ \ \ \ \ \ \ \ \ \ \  and to match it to the red end of Ly$\alpha$ satellites from {H--H} collisions \citep{allard:NewQuasi-H2-Lya}.  $^{\rm c}$Completeness

    \ \ \ \ \ \ \ \ \ \ \ \ \  is important for non-LTE calculations; neutral atoms have highest priority.
\end{table}

\subsection{Some Recent and Pending Calculations}

A new high-temperature (up to 1500\,K) line list for acetylene (HCCH or C$_2$H$_2$) has been published by \citet{lyulin:HiT-C2H2lineparms} with several experimental spectra cited. \citet{chubb:C2H2ExpCompilation}
performed a critical compilation of measured spectra and derived rovibrational parameters for acetylene.

Calculations of collision-induced absorption has been carried out by \citet{karman:N2-N2-CiA} at 50--330\,$\mu$m and \citet{hartmann:N2-N2-CiA} at 2.05--2.3\,$\mu$m, but only for room temperature and below. Experimental spectra are cited in both papers.

The ExoMol team \citep[][Tennyson, this volume]{yurchenko:ExoMol} is very productive, and has already addressed a great number of requests from the community, notably HCN/HNC \citep{barber:HCN-HNC} and NH$_3$ \citep{yurchenko:NH3} with updates expected early 2018. In addition, an ExoMol calculation of C$_2$H$_2$ is slated to be published by the first half of 2018, and C$_3$ should be completed later on that same year.

\acknowledgements RT acknowledges funding from NASA grant NNX15AB24G. Funding for the Stellar Astrophysics Centre is provided by The Danish National Research Foundation (Grant DNRF106).

%\bibliography{bibs/conv,bibs/opac,bibs/eos,bibs/starmod,bibs/seism,bibs/obs,bibs/math,bibs/staratm}  % For BibTex

%\input{Trampedach_OpacWishList.bbl}

\end{document}